# Advances in Atomic Resolution In Situ Environmental Transmission Electron Microscopy and 1Å Aberration Corrected In Situ Electron Microscopy

PRATIBHA L. GAI[1,2*] AND EDWARD D. BOYES[2,3]
[1]Department of Chemistry, University of York, Nanocentre, York YO10 5DD, United Kingdom
[2]Department ofPhysics, University of York, Nanocentre, York YO10 5DD, United Kingdom
[3]Department ofElectronics, University of York, Nanocentre, York YO10 5DD, United Kingdom

ABSTRACT Advances in atomic resolution in situ environmental transmission electron microscopy for direct probing of gas-solid reactions, including at very high temperatures ( 2000 °C) are described. In addition, recent developments of dynamic real time in situ studies at the Angstrom level using a hot stage in an aberration corrected environment are presented. In situ data from Pt/Pd nanoparticles on carbon with the corresponding FFT/optical diffractogram illustrate an achieved resolution of 0.11 nm at 500 °C and higher in a double aberration corrected TEM/STEM instrument employing a wider gap objective pole piece. The new results open up opportunities for dynamic studies of materials in an aberration corrected environment. Microsc. Res. Tech. 72:153–164, 2009.

## INTRODUCTION

Many dynamic processes occur in nature, science or industry in reaction environments. Therefore, dynamic in situ data related to the real world need to be obtained under controlled conditions of gas/vapor/liquid environment and temperature. The direct observation of nanostructural evolution, metastable states, and mechanisms of reactions under dynamic reaction conditions is a powerful structural tool in the chemical and materials sciences, particularly in heterogeneous catalysis and solid state defect phenomena (Boyes and Gai, 1997; Gai, 1981, 1992; Gai et al., 1990, 1995), semiconductor studies (Sinclair et al., 1981), and metallurgy (Butler and Hale, 1981). In situ E(S)TEM under controlled reaction conditions provides dynamic information on processes, performance critical defect structures and sub-surface diffusion of catalytic species, which can not be obtained directly and readily by other techniques.

In a number of earlier environmental transmission electron microscope (ETEM) studies using modest resolution reported in the literature, different approaches were employed to contain gas (or liquid) environments inside a TEM's vacuum column. In early ETEM systems, environmental cells (ECELLs or micro/nanoreactors) were designed to be inserted between the objective lens pole-pieces of the electron microscope (EM) for in situ studies, necessitating the frequent opening and rebuilding of the EM. Typically pairs of apertures are added on either side of the sample with differential pumping lines attached between them. Reviews of early work in the materials and chemical sciences (Butler and Hale, 1981; Crozier et al., 1998; Gai, 1981, 1992), and references therein, illustrate the significant scientific impact of ETEM. However to better understand atomic scale processes in reacting materials, atomic resolution-ETEM is required. The development of the atomic resolution in situ-ETEM for studies of gas-solid reactions at the atomic level and applications including at high temperatures (at 2000°C), are reported below. We then describe recent advances in in situ aberration corrected electron microscopy in a 1Å aberration corrected environment.

## EXPERIMENTAL PROCEDURES
### Pioneering Development of Atomic-Resolution ETEM for In Situ Studies of Gas-Solid Reactions in Controlled Environments: A Powerful Technique for Materials Research

Recently, atomic resolution-ETEM has been developed by the authors Gai and Boyes (Boyes and Gai, 1997, Gai, 1997; Gai et al., 1995) for probing of gas-solid reactions directly at the atomic level under controlled reaction conditions. Temperature, time and pressure resolved studies with high precision are possible. Briefly, the whole EM column in a 300 kV CM30T high resolution (S)TEM has been extensively modified with the introduction of a fully integrated and permanently mounted ECELL system for atomic resolution studies of dynamic gas-solid reactions and for in situ nanosynthesis (Boyes and Gai, 1997; Gai, 1997, 1998). Highlights of this development (Boyes and Gai, 1997; Gai, 1997), shown in Figure 1a, include a novel ETEM design with objective lens pole-pieces incorporating radial holes for the first stage of differential pumping: in addition, the controlled environment ECELL volume is the regular sample chamber of the ETEM and is thus integral to the ETEM. It is separated from the rest of the column by the apertures in each pole-piece and by the addition of a gate valve, in the line to the regular ion-getter pump (IGP) at the rear of the column. Differential pumping systems are connected between the apertures using molecular drag pumps (MDP) or turbo-molecular pumps (TMP). This permits high gas pressures in the sample region, while maintaining high vacuum conditions in the rest of the ETEM.

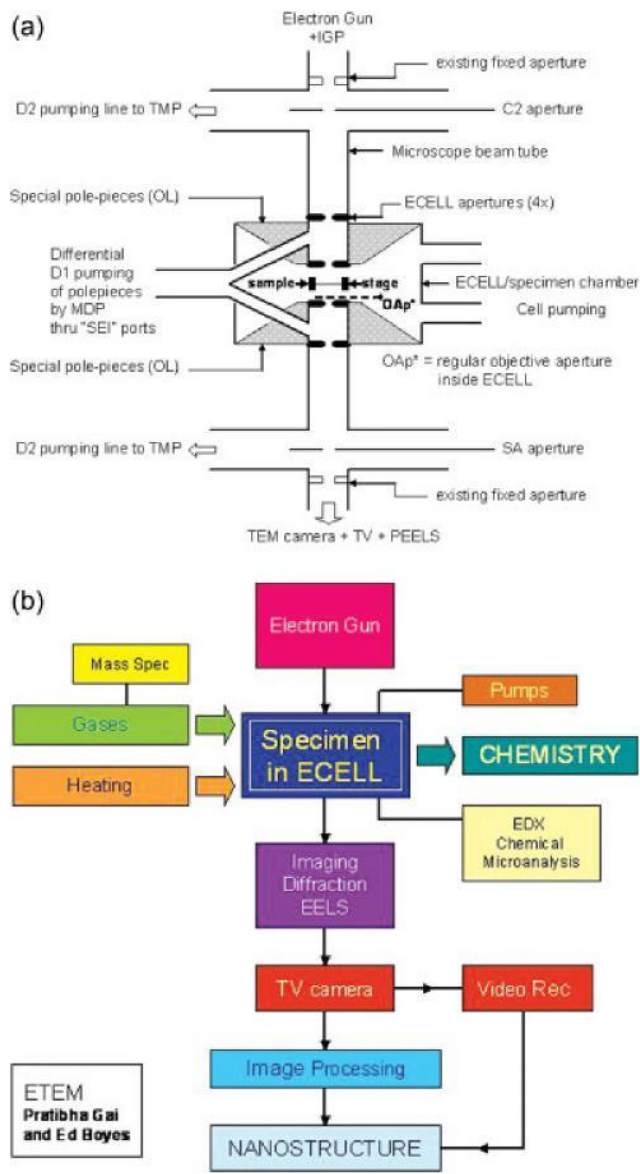

Fig. 1. a: A schematic of the basic geometry of the aperture system in the pioneering development of atomic resolution in situ ETEM by Gai and Boyes (Gai et al., 1995; Boyes and Gai, 1997) to probe gas-solid reactions at the atomic level. Radial holes through objective polepieces (OL) are for gas lines at D. The OL polepiece are above and below the specimen holder and the lower OL polepiece is indicated. Gas inlet, the first stage of differential pumping lines (D1) between the environmental cell (ECELL) apertures, condenser aperture, a second stage of pumping (D2) at the condenser lens, selected area (SA) diffraction aperture, parallel electron energy loss spectroscopy (PEELS) and TEM camera vacuum are indicated. The novel ETEM design of Gai and Boyes has been adopted by commercial TEM manufacturers (FEI) and the development has been highlighted by the American Chemical Society's C&E News(1995, 2002). b: Schematic of the accessories to in situ ETEM for probing gas-solid reactions. [Color figure can be viewed in the online issue, which is available at www.interscience.wiley.com.]

Dynamic atomic resolution imaging and electron diffraction are complemented by chemical analysis capability by incorporating a GIF-PEELS (Gatan imaging filter-parallel electron energy loss spectroscopy) and scanning transmission EM (STEM) attachment in the atomic-resolution ETEM (Boyes and Gai, 1997). A conventional reactor-type gas manifold system enables inlet of flowing gases into the ETEM, and a sample stage with a furnace (hot stage) allows samples to be heated up to 1000°C (and in special cases, higher, as described in the following sections). For dynamic atomic resolution, a few mbar of gas pressures are routinely used in the ECELL (higher gas pressures are possible with some loss of resolution due to multiple scattering of electrons through thicker gas layers). A mass spectrometer is added for gas analysis. Figure 1b shows a schematic of some of the accessories to the ETEM. Very low electron dose techniques are used for in situ experiments. In situ data are checked in a parallel calibration (blank) experiment with the electron beam switched off and the sample exposed to the beam only to record the reaction end point (Gai, 1992). This ensures a completely noninvasive characterization. A video system connected to the ETEM facilitates digitally processed recording of dynamic events in real-time, with a routine time resolution of 1/30 s. Some important features of this development are shown in Figure 1. The development was highlighted by the American Chemical Society's C&E News (Haggin, 1995; Jacoby, 2002). The novel design of the atomic resolution-ETEM by Gai and Boyes (Boyes and Gai, 1997; Gai, 1997; Gai et al., 1995) has been adopted by TEM manufacturer FEI for commercial production and later versions of the Gai and Boyes' in situ ETEM instrument, (which include, the CM 200–300 series, Tecnai and Titan (including the aberration corrected versions) have been installed in laboratories around the world (e.g., Materials Research Society Bulletins: Ferreira et al., 2008; Gai et al.,2007a, 2008; Hansen et al., 2001; Sharma et al., 2007).

Under carefully simulated conditions, data from in situ ETEM can be directly related to structure-activity relationships in technological processes. In catalysis, this correlation is crucial in optimizing the synthesis and development of improved catalysts to ensure the catalyst life of a few or more years, as such long periods are typically used in catalytic technology. Several key selective hydrocarbon oxidation experiments in the ETEM have shown that gas pressures of a few mbar on the catalyst surface are adequate for insights into gas-solid reactions and for the technological correlation and that the catalyst surface coverage is more important than higher pressures around the sample (Gai, 2007; Gai and Boyes, 2003). Atomic resolution-in situ ETEM images of molecular sieves in high gas pressures of 0.5 bar of hydrogen have also been reported (Thomas et al., 2001). Because of the small amounts of solid reactant in the microscope sample, measurement of reaction products are performed on larger samples in a microreactor operating under similar conditions and used for nanostructural correlation.

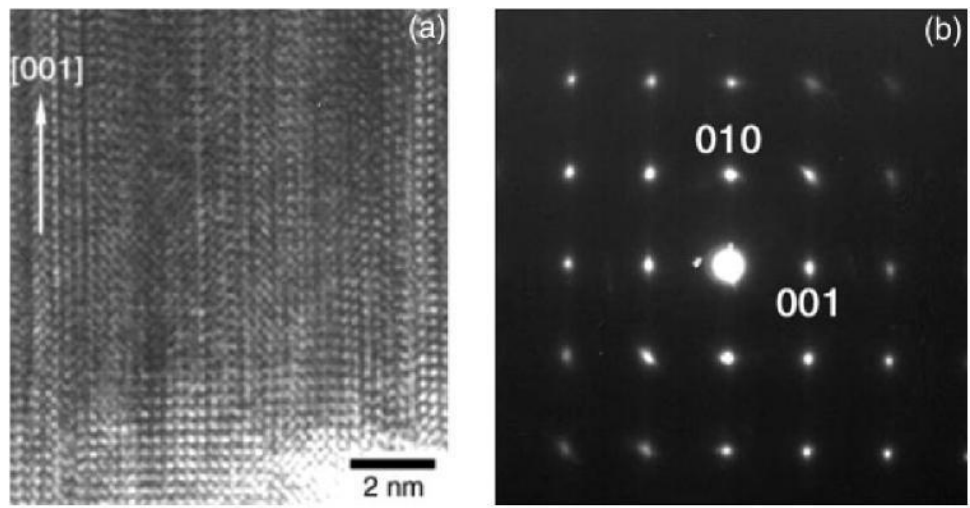

Fig. 2. Atomic resolution image of (100) hexagonal WC. a: Atomic scale twin boundary structures along the [001] direction with lattice modulations along [010]. b: Electron diffraction pattern of (100) WC. 001 and 010 reflections are indicated.

### Dynamic Atomic Scale Twinning Transformations and the Formation of Carbon Nanostrutures in Tungsten Carbide (WC)

Active transition metal carbides are used in heterogeneous catalysis, fuel cells and nanoelectronics technologies (Oyama, 1996). The dynamical behavior of hexagonal tungsten carbides (WC) employed in the form of powders as heterogeneous catalysts for the hydrogenation reaction of benzene, methanation and in fuel cell electrodes is therefore of great importance in these technologically important reactions (Gai et al., 2007b). The current understanding of the operation of WC however, is based on ex situ chemical studies which have considered only a perfect WC structure, which is uncommon, as WC is an inherently imperfect structure (Gmelin, 1991; Massalski, 1986). This has hindered the understanding of the performance of WC. Direct atomic-level experimental results from in situ ETEM in reducing atmospheres in real time are therefore key to better understand the performance of WC in the technologically relevant reactions.

Spectrographically pure WC powders were dispersed in alcohol and supported on finely meshed nickel or titanium grids for ETEM studies. Figure 2a is an atomic resolution image showing extensive twinning observed in complex WC particles in (100) crystallographic orientation. WC is hexagonal (space group P-6m2), with a 5 0.2906 nm and c 5 0.2838 nm (Massalski, 1986; Oyama, 1996). In the hexagonal lattice, the four axes (hkil) of Miller-Bravias indices are abbreviated to (hkl), where, i 5 2(h 1 k). Thus, (10-10) is the (100) crystallographic orientation. The (001) lattice planar spacings of 0.28 nm and the (010) lattice spacings of 0.25 nm are clearly resolved in the image. The lattice spacings ($d_{hkl}$) in the hexagonal structure were obtained by using the formula (Hirsch et al., 1985):

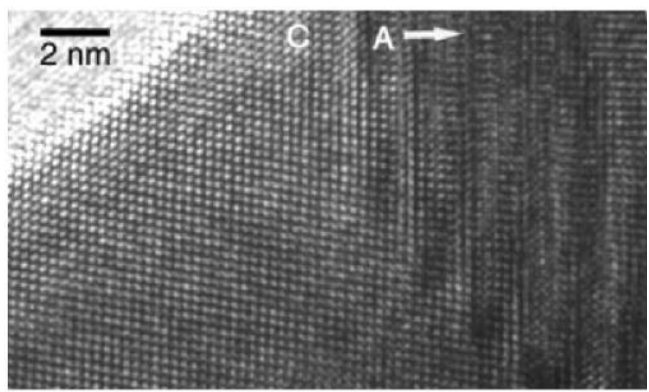

Fig. 3. Atomic resolution image of a complex bicrystalline WC grain, showing the presence of twin defects in (100) but their absence in (001) orientation, denoted by A and C, respectively.

$$\left(1 = d_{hkl}\right) = \left(4 = 3\right)\left(h^2 + hk + k^2\right) = a^2 + \left(1 = c\right)^2$$

Figure 2a shows twin boundaries along the [001] direction and lattice modulations along [010]. Figure 2b shows an electron diffraction pattern of WC in (100), with the main (010) and (001) reflections indicated. To exclude the possibility of three-dimensional configuration for the atomic twins, we imaged complex bicrystalline WC grains. Figure 3 is an atomic resolution image of a bicrystalline WC grain in (100) and (001) crystal projections denoted by A and C, respectively. The image clearly demonstrates that the twin structures are present in (100), but are absent in (001) orientation, thus establishing the planar nature of the twins.

In situ ETEM studies of the same thin area of (100) WC sample were carried out in 20% $H_2$/He at a pressure of about 3 mbar and the results are shown in Figure 4. Figure 4a shows atomic twin defects at room temperature (RT). The twins are along [001] with lattice modulations along [010]. Figure 4c shows the corresponding electron diffraction (ED). Figure 4b and 4d show the dynamic image and ED, respectively, of the same area at 450°C, reacted for 15 min. Figure 4c demonstrates the elimination of almost all of the twin structures and the transformation to W nano-metal and amorphous carbon. The presence of the metal was confirmed by the emergence of (100) reflections due the cubic a-W (with a 5 0.32 nm) in the parent crystal and by compositional analysis. At higher temperatures of 850–900°C, a variety of carbon nanostructures, including graphite, carbon nanotubes as well as amorphous carbon, are observed in thin regions of the samples.

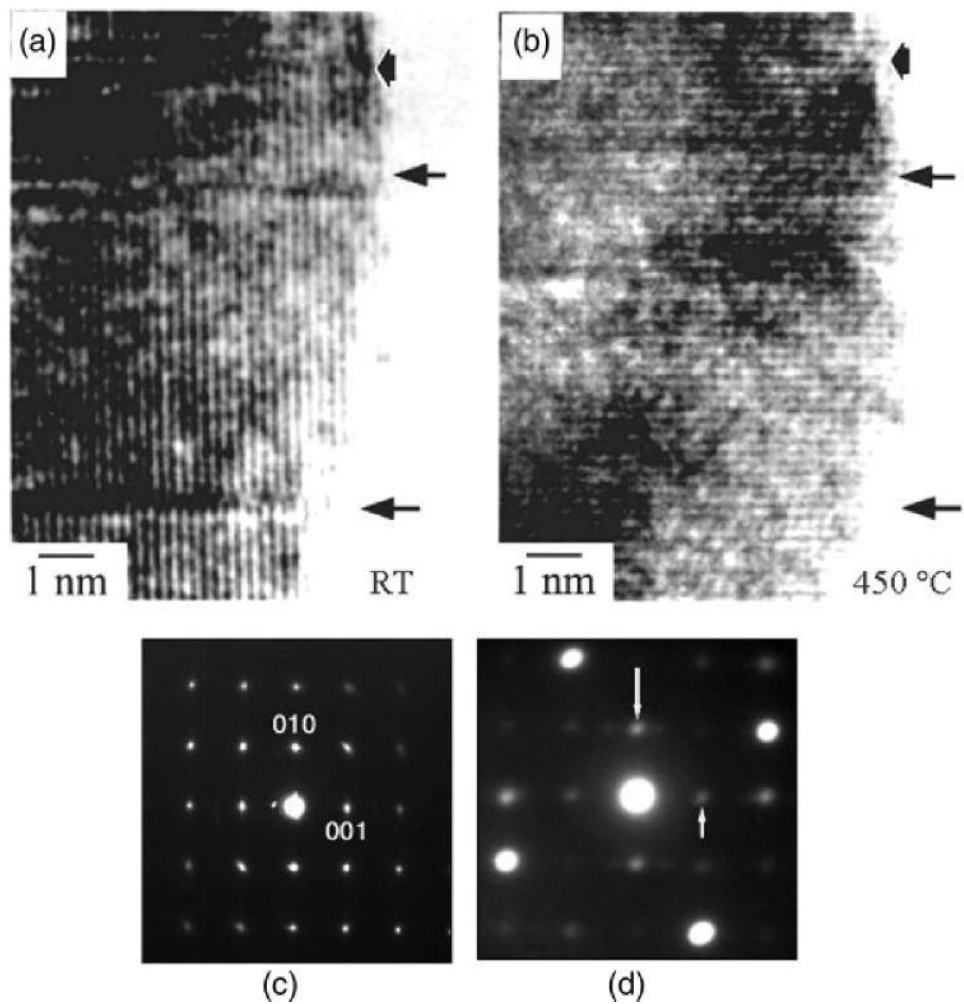

Fig. 4. In situ ETEM studies of real-time reduction of the same area of WC powder sample in hydrogen/He gas: (a) room temperature (RT); (c) the corresponding electron diffraction; (b) dynamic reaction at 450°C, indicating the elimination of almost all of the twins; (d) dynamic electron diffraction pattern of (b) indicating the emergence of (100) cubic tungsten reflections (arrowed).

Figure 5a shows the structure of WC in (100) crystal projection. The carbide adopts a hexagonal crystal structure with nonmetallic carbon atoms in the interstitial positions between the metal atoms. Twin structures are believed to form in carbide grains during their synthesis accommodating the disorder and reorder of the crystal structure during the grain growth to minimize the grain boundary energy (Cottrell, 1971). Based on our observations in Figure 4 we propose a possible structural and compositional model for the atomic twin boundary structure along (001). This is shown in the projection of one layer of the WC structure on (100) in Figure 5b with carbon atoms gathered in the boundaries. We believe that in reducing hydrogen environments, WC containing atomic twin structures, is thermodynamically unstable and there is a strong driving force for the reduction of the carbide. Relatively loosely bound carbon atoms in the twin structures interact with the $H_2$ gas and diffuse out of the surface. This is accompanied by readjustments of atomic positions to produce Wand carbon structures.

On the basis of these newly discovered dynamic atomic scale interactions between a reactive tungsten carbide and reductive gas environments, we conclude that the reaction kinetics of the carbide are altered by diffusion processes along atomic twin boundaries leading to possible loss of some carbon atoms. This has important implications in understanding the chemical reactivity of the carbide in commercially important heterogeneous catalysis and fuel cell technologies. By real-time monitoring in hydrogen, we have demonstrated that the atomic scale twin structures play a direct and key role in the chemical reaction. We further speculate that similar atomic interfacial transformations should be possible in other transition metal carbides as well as nitrides.

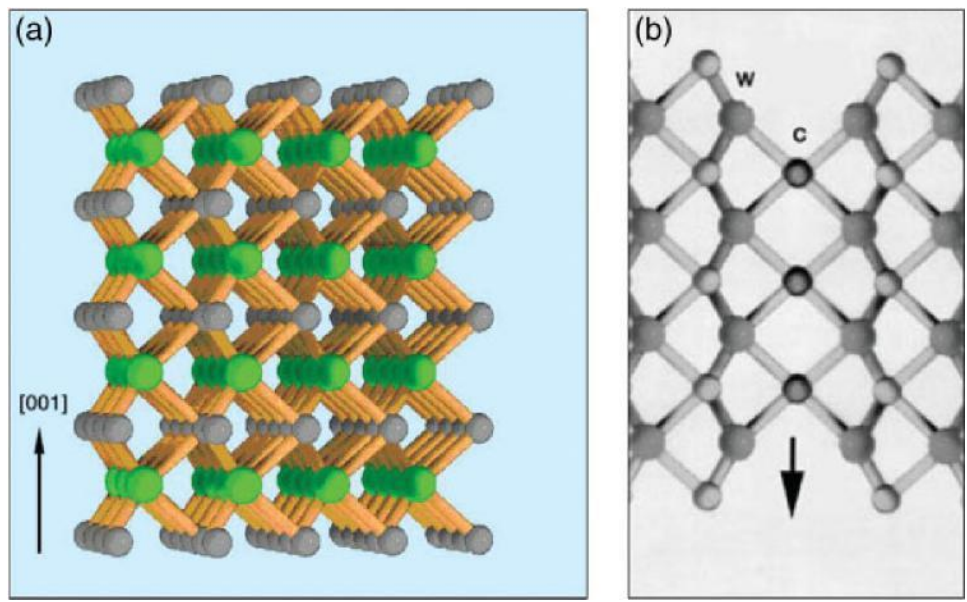

Fig. 5. a: Structural schematic of hexagonal WC in (100) orientation (a ≅ 0.2906 nm and c ≅ 0.2838 nm). W atoms are shown in green. Carbon atoms (in black) occupy interstitial positions. b: Projection of one layer of the WC structure on (100), showing a possible twin boundary structural model along (001) (arrowed), consistent with the experimental observations. Carbon atoms gathered in the defects react with hydrogen and diffuse out. This is accompanied by readjustments of the atoms to create the metal structure. [Color figure can be viewed in the online issue, which is available at www.interscience.wiley.com.]

### Development of a Very High Temperature (2000°C) Stage for Atomic Resolution—In Situ ETEM

Some important applications in the chemical and materials sciences require very high temperatures under gaseous atmospheres. For example, the formation of ceramics, carbon nanotubes, the conversion of carbon based systems, and phase transformations in white pigments require temperatures in the region of 1100–2000°C in reactive gases. Commercial holders for ETEM are limited to about 1000°C (Boyes and Gai, 1997; Crozier et al., 1998; Gai, 1992; Gai et al., 1995, 2007a). Therefore, there is the need to develop a very high temperature hot stage (about 2000°C) for applications in the atomic resolution-in situ ETEM.

We have designed a very high temperature holder to meet these demands. The environmental very high temperature holder (EVHT) design is based on the early work in vacuum by Kamino and Saka (1993). We have used stainless steel and brass components for the holder and tungsten (W) filaments (Gai and Boyes, 2008a). The diameter of the filament was selected after testing filaments with different sizes. The holder was extensively safety tested including for X-ray radiation and it has electrostatic screening. Experiments were carefully performed using the EVHT in the ETEM.

Figures 6a–6c show the holder and the tip at different magnifications. We calibrated temperatures using an optical pyrometer in a test rig. Temperatures were recorded from 800°C up to 2200°C and up to 2300°C in a few cases. Temperature versus milliamp (mA) calibration data with optical pyrometer readings through glass are shown in Table 1 in Figure 6d. In the test rig, experiments were performed on the EVHT holder in different atmospheres, namely nitrogen, hydrogen and hydrocarbons.

Samples of cellulose were then attached to the filament and the holder was transferred to the ETEM. Nitrogen gas was introduced at a pressure of 1 mbar and reactions were carried out on thin fragments of the sample with reaction times of a few minutes. Initial studies indicate the transformation of cellulose into graphitic containing structure at about 2000°C (Figs. 7a and 7b). Cellulose is an organic compound containing hydroxyl groups which form H-bonds with oxygen molecules on different chains and part of the structure is shown in Figure 7c. We believe that the transformation is associated with the removal of the combined water, OH groups, and protons.

### In Situ Probing of Sintering of Supported Nanoparticles

A variety of industrial catalytic processes employ metal nanoparticles on porous inorganic supports. Commonly used supports are carbon, or ceramic oxides such as silica or alumina. The catalysts contain a high dispersion of nanoscopic metal particles on these supports. This is to maximize the surface area with a minimum amount of metal for catalytic reactions (Thomas and Gai, 2004). There has been considerable work in the literature on the structure of very small particles. In a significant number of nanoparticle preparations/processing, a sizable portion of nanoparticles exhibit noncrystallographic structures such as multiply twinned particles (MTPs) (Anderson, 1972; Gai and Boyes, 2003). Figure 8 shows an example of a gold multiply twinned nanopar-ticle on germanium support at room temperature. There is considerable interest in the structure and the role of such defects in ultra small nanoparticles (in the 1–3 nm range) thought to be active species in catalytic reactions and type of the noble metal in processing.

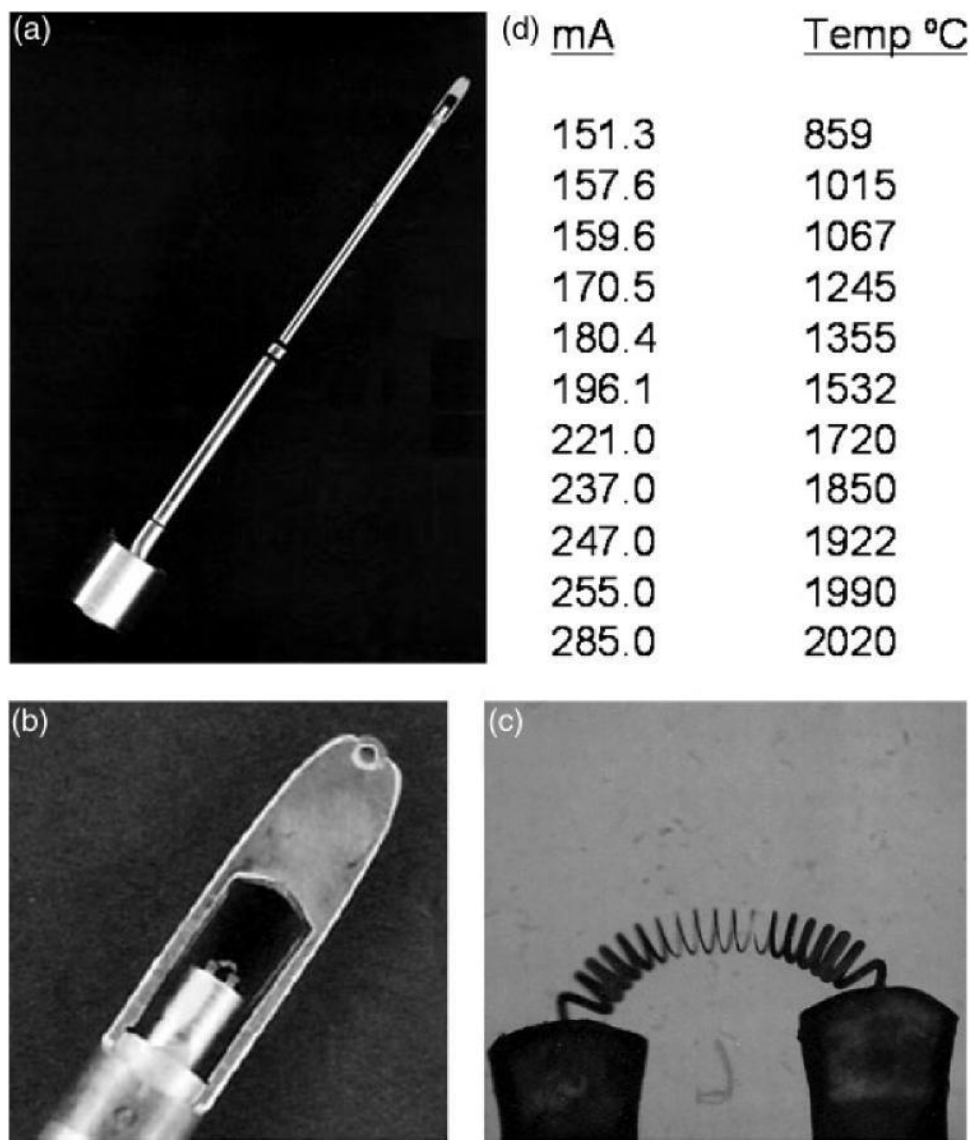

Fig. 6. (a) Very high temperature (2000°C) stage (holder) for ETEM; (b) and (c) show the holder tip at different magnifications. (d) Table 1: Calibration data of temperature versus milliamp (mA) in 1 mbar $N_2$.

To obtain insights into the dynamics of ultra-small particles we have developed in situ aberration corrected electron microscopy at the Angstrom scale described in the following paragraphs.

### New Opportunities With Aberration Corrected Electron Microscopy: Dynamic In Situ Nanoparticle Sintering Experiments in a 1A Aberration Corrected Environment

Nanoparticles of Pt, Pd, and Au with only a few nanometres (nm) in sizes are important in technological chemical processes and sintering impacts their performance. Bimetallic systems such as Pt-Pd offer better performance compared with monometallic systems in many applications such as energy conversion, catalysis, nanoelectronics, chemical and biological sensing and photonics. There have been reports of the high reactivity of gold nanoparticles including synthesis and processing in vacuo (e.g., Daniel and Astruc, 2004; Yen et al., 2007) with operating temperatures up to 500°C. However, understanding of the sintering behavior in processing of supported ultra small particles (1 nm to a few nm) at the atomic or Angstrom level is lacking and only larger particles, nanowires, nanorods and nanoplates have been reported in the literature (Gall et al., 2005; Kan et al., 2006; Mohamed et al., 1999). Thus far, aberration corrected electron microscopy studies have been reported only at room temperature in the literature. In situ methods are particularly informative in catalysis and studies of materials as described in the previous sections. To be useful they need to be conducted under controlled conditions of temperature, and where appropriate environment; and always with minimal, mechanistically non-invasive, influence of the electron beam. This means that the microscope must be able to accept a hot stage, if heating is the driving force for change being investigated; with all other conditions held constant.

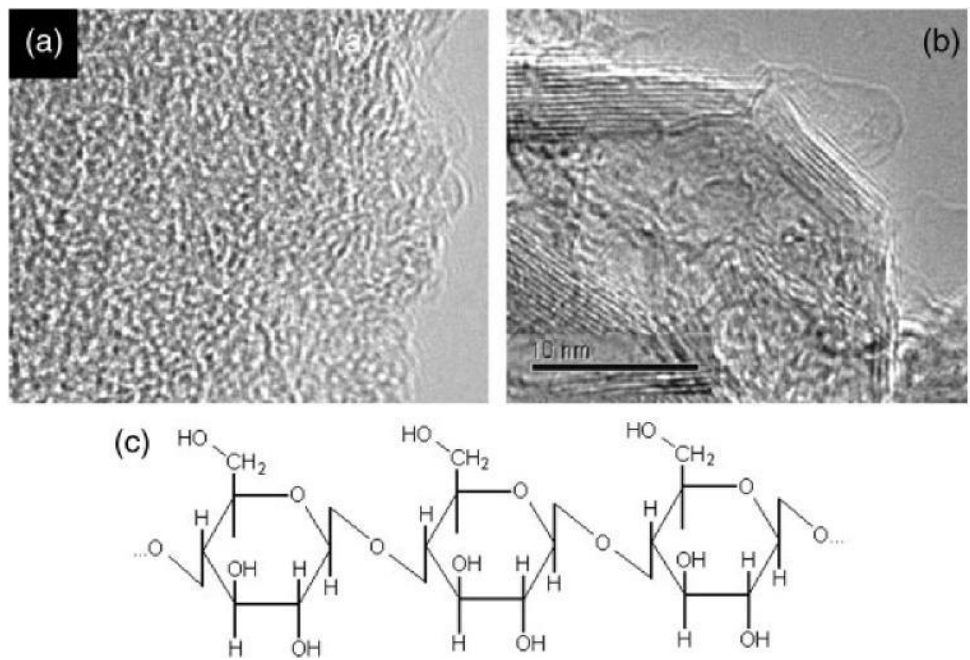

Fig. 7. In situ ETEM studies of transformation of cellulose in 1 mbar nitrogen. a: Cellulose precursor at room temperature (RT). b: Graphitic structure at 2000°C. c: Schematic of a part of the cellulose structure. Some atoms are indicated.

Aberration correction is particularly beneficial in dynamic in situ experiments because there is rarely, if ever, the opportunity to record from a scene which may be continuously changing, a systematic through focal series of images for subsequent data reconstruction. In these applications, it is necessary to extract the maximum possible information from each single image in a continuously changing sequence. It is also desirable where possible to limit the electron dose exposure to ensure minimally invasive conditions and to avoid secondary effects such as contamination. Based on these considerations we proposed aberration corrected dynamic in situ electron microscopy (Gai and Boyes, 2005, 2006, 2007).

The need to accommodate special specimen holders in an aberration corrected machine has been one of the more important criteria driving the specification of the JEOL 2200FS double aberration corrected (2AC) FEG TEM/STEM operating at 200 kV in the new Nanocenter at the University of York. Since aligning the sample into a zone axis orientation is a prerequisite for atomic resolution electron microscopy of crystalline materials, an increased specimen tilt range was also desired. Both these conditions benefit from the larger gap (HRP) objective lens polepiece, and in this case the Cs (C3) aberration correctors (Haider et al., 1998) on both the STEM probe forming and TEM image sides of the instrument were used to provide the desired expanded specimen geometry with minimal effect on the 1Å imaging performance of the system. The effect of the increased polepiece gap on Cs in this range is much more evident (32) than on Cc (31.2); and of course Cs is now corrected.

## Procedures

We sputtered Pt-Pd nanoparticles (with 1:1 ratio) on carbon, which are of interest in catalysis and fuel cell technology, to an average thickness of 2.5 nm with particle sizes of <1–2 nms (Wright, 2008) and gold nanoparticles on carbon, following methods similar to those reported (Gai and Smith, 1990). To observe changes in the nanostructure as a function of operating temperature under controlled in situ conditions in an aberration corrected environment, we employed the following procedures.

The advantages of the configuration we have adopted, include promoting a contrast transfer function (CTF) extending to higher spatial frequencies and resolution in the data; allowing image recording at close to zero defocus, including to strengthen and simplify interpretation of information at internal interfaces and external nanoparticle surfaces; analyzing small (<2 nm) nanoparticles and clusters on supports, using high resolution TEM as well as HR STEM; facilitating HAADF STEM and extending HAADF STEM resolution to 1Å (0.1 nm) and below (Allen et al., 2003; Batson et al., 2002). As well as benefiting from improved resolution at 1Å and below, it becomes important to be able to set the conditions to avoid the previously intrusive CTF and defocus sensitive oscillations in image contrast in the spatial resolution range from 1Å to 3Å. This is where the atomic neighborhoods in crystals lie in low index projections and it is especially important in studying the surfaces of ultra-small nanoparticles. It is a key topic of interest in nanoparticle studies, e.g., in considering the possible origin of sintering mechanisms important in heterogeneous catalyst design including processing/calcination in vacuo and understanding deactivation. We believe is an aspect of nanotechnology to which electron microscopy is uniquely well qualified to contribute. Calibration experiments were performed according to procedures described in "Pioneering Development of Atomic-Resolution ETEM for In Situ Studies of Gas-Solid Reactions in Controlled Environments: A Powerful Technique for Materials Research" section. The calibration ("blank") experimental procedure (without the beam) was employed for longer reaction times of a few hours.

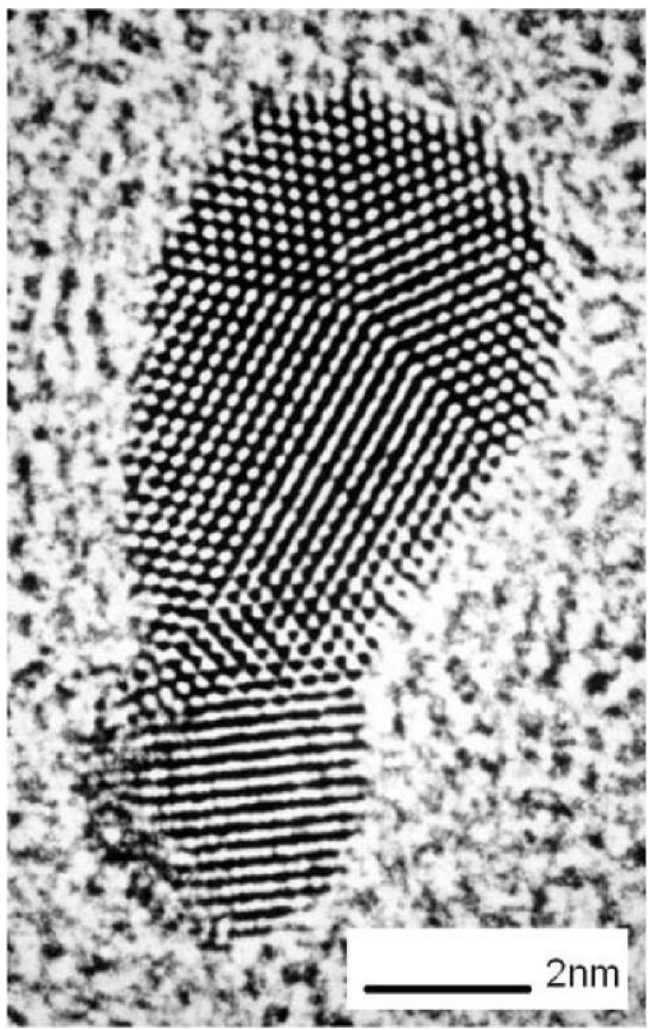

Fig. 8. Multiply twinned gold nanoparticle on germanium film at 20°C.

## RESULTS AND DISCUSSION

We show in both theory (www.maxsidorov.com)#, with contrasting CTFs shown in Figure 9, and in practice (Fig. 10), it is possible with Cs aberration correction to combine demonstrated spatial resolution around 1A with the larger gap (HRP) lens polepiece with 4.3 mm gap (Gai and Boyes, 2008b) needed to accommodate a standard hot stage (Gatan model 628); and with it in operation (using Gatan Digital Micrograph). This is an example of using aberration correction to combine the limited added space required for in situ experiments with a high level of imaging performance with which such facilities were previously incompatible, and thereby to extend considerably application specific and relevant TEM and STEM experimental capabilities.

The system is in practice stability limited (CTF envelope terms) and some of the practical steps necessary to deliver this powerful combination of capabilities will be covered elsewhere with further examples of the new tool in action. These considerations quickly set a limit to how far the lens gap can be stretched without beginning to compromise performance too seriously; taking into consideration realistically attainable stabilities in internal electronics and mechanics, and in key external environmental factors. Our instrument is further con­figured with three turbomolecular pumps as the main column vacuum system to be tolerant to outgassing hot samples, and as a key foundation for controlled gas environment experiments. The "stretched"instrument configuration is considered to be a prerequisite for fur­ther in situ developments adding additional facilities and capabilities (Boyes and Gai, 1997; Gai et al., 1995, 2007a) to the exciting and still relatively new generation of aberration corrected TEM and STEM machines. In our case both capabilities are combined on the remotely controlled JEOL 2200 FS platform.

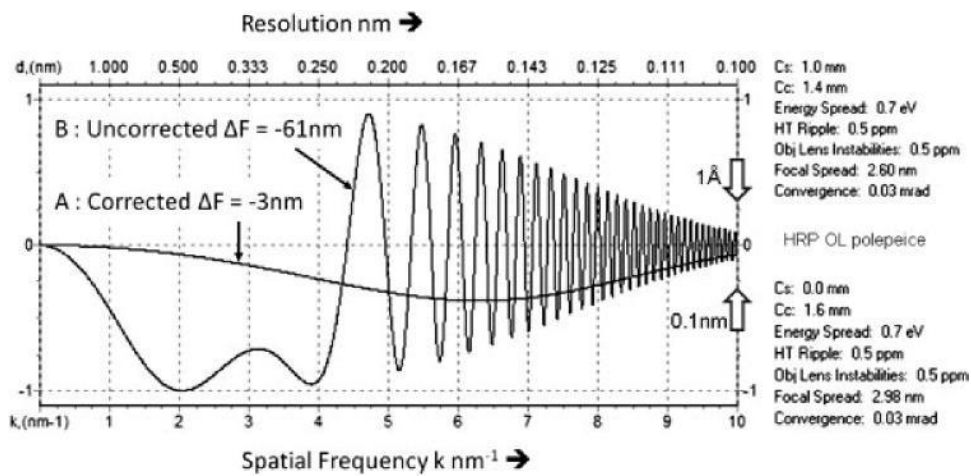

Fig. 9. Calculated contrast transfer functions (CTFs) for uncorrected (A) and aberration corrected (B) imaging conditions of the HRP version of the double aberration corrected JEOL 2200FS FEG TEM/ STEM STEM at The University of York (UK).

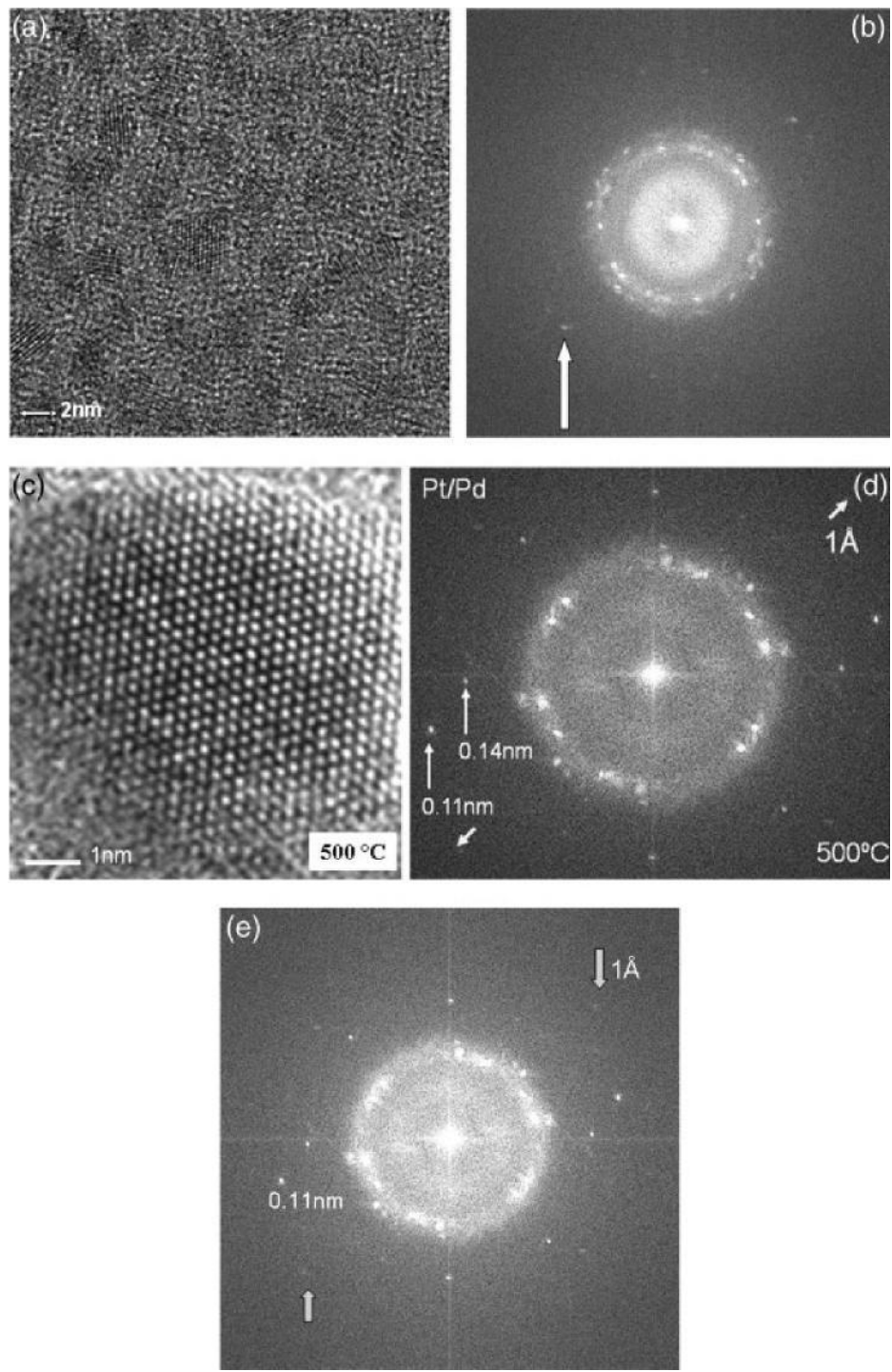

Fig. 10. Dynamic real time in situ studies in the JEOL 2200FS double-aberration corrected (2AC) TEM/STEM (2AC TEM/STEM) at the University of York. (a) Pt-Pd nanoparticles on carbon support at room temperature (RT) (b) the corresponding FFT/optical diffracto­gram with 0.11 nm resolution. The particles are in the size range <1– 2 nm with the carbon support contributing strongly to the diffraction. (c) In situ hot stage image of a selected Pt-Pd nanoparticle on carbon at 500°C, (d) the FFT (optical diffractogram (OD) equivalent) showing the distinctive image resolution of 0.11 nm, using a Gatan 628 hot stage. (e) FFT/OD equivalent at 500°C with 0.1 nm resolution.

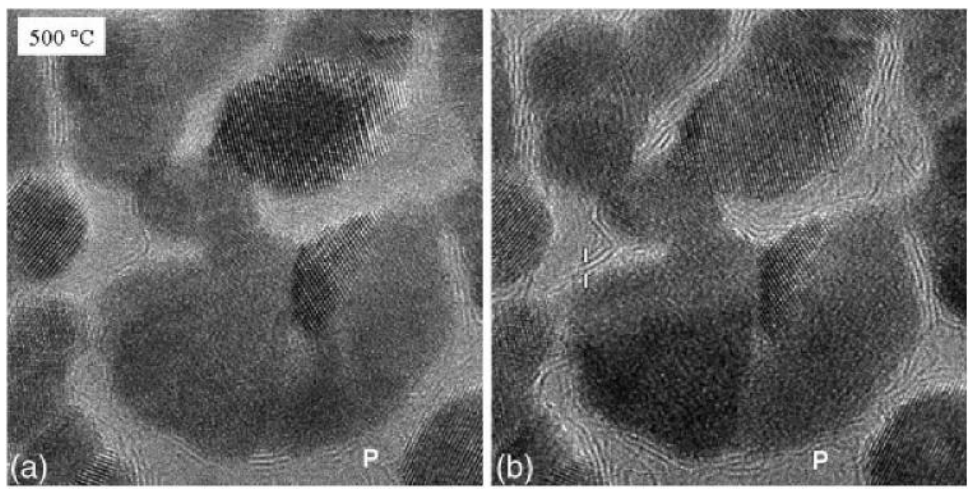

Fig. 11. Dynamic in situ studies of sintering as a function of time (the same area near particle P) at 500°C in the 2AC TEM/STEM: (a) after 3 h indicating onset of graphitization (b) after 4 h with increased graphitization and striking changes in the support structure. Graphite spacing of 0.34 nm is indicated.

TEM correction montage procedures (Zemlin tableaux) and image resolution to 1A have been maintained with the hot stage, including with power connected and the samples held at elevated temperatures; so far to 900°C. Figure 10a shows an image of Pt-Pd nanoparticles on carbon support at room temperature (RT) and Figure 10b shows the corresponding FFT/optical diffractogram with 0.11 nm resolution. The particles are in the size range <1–2 nm with the carbon support contributing strongly to the diffraction. Figures 10c–10e illustrate the dynamic in situ data from the Pt/Pd nanopaticles on carbon after particle coarsening, sintering and grain growth in the in situ heating experiments after 30 min, with the corresponding FFT/optical diffractogram illustrating achieved resolution of 0.11 nm at 500°C with the Gatan 628 hot stage operating in the JEOL 2200 FS 2AC TEM/STEM. The changes due to the sintering and the formation of larger nanopaticles can be seen in Figure 10d. Figures 11a and 11b show the same particle (for example at P) at 500°C as a function of time after 3 h and 4 h respectively. The onset of graphitization was observed after 3 h and increased graphitization after 4 h as shown in Figure 11b, revealing striking changes in the support nanostructure. The graphitization was observed to inhibit further the sintering of the nanoparticles and was found to increase with temperature and time. There was little evidence of electron beam effects contributing to the Pt/Pd nanostructure development and sintering under TEM microscope operating conditions and the processes seemed to be driven predominantly by the thermal input from the controlled and calibrated hot stage. However, this was not the case with some forms of gold nanoparticles on a similar commodity carbon support film with some modifications around the particle, and considerable care in operation was then needed to limit beam effects. These effects were also reported in early STEM experiments (Isaacson et al., 1977) as well as more recently (Batson et al., 2002). The high resolution data indicate that with neither of our samples were any contamination effects seen.

Figure 12a illustrates three gold nanoparticles at RT and their sintering into a larger agglomerate at 500°C (shown in Fig. 12b), with alignment of crystal planes as this occurs. In the process, the center of gravity of one of the particles has migrated and the structure has rearranged quite considerably, with parallel alignment of (111) planes in adjacent nanoparticles. However, some remote segments of other particles appear to be largely unaffected in the process retaining the internal crystallography and external (111) surface terrace features. We believe that sintering occurs due to the initial reduction of the surface disorder in (111), the driving force for sintering being the reduction in the surface area and surface energy of the disordered (111) facets. This appears to be applicable to other nanoparticles including MTPs where surface disorder is present. Different MTP structures have been described including the decahedron as being one of the basic structures consisting of five tetrahedral units bound by (111) twin planes (Marks and Smith, 1981). The power of aberration correction in enabling high resolution data recording at close to zero defocus is illustrated in Figure 12c from a performance critical size range (2–4 nm) of gold nanoparticle at 25°C, showing image detail of (111) surface terraces important in sintering. Very evident in the image is the avoidance of most of the image delocalization effects typically seen in such images recorded without Cs aberration correction under conditions far from zero defocus in pursuit of image contrast. Several proposals exist about active sites of nano-gold on different supports (Bondize et al., 1999; Varykhalov et al., 2008): these include the possibility of the presence of neutral gold atoms on the surface of thin film of gold completely coating titania support although atomic resolution studies of the

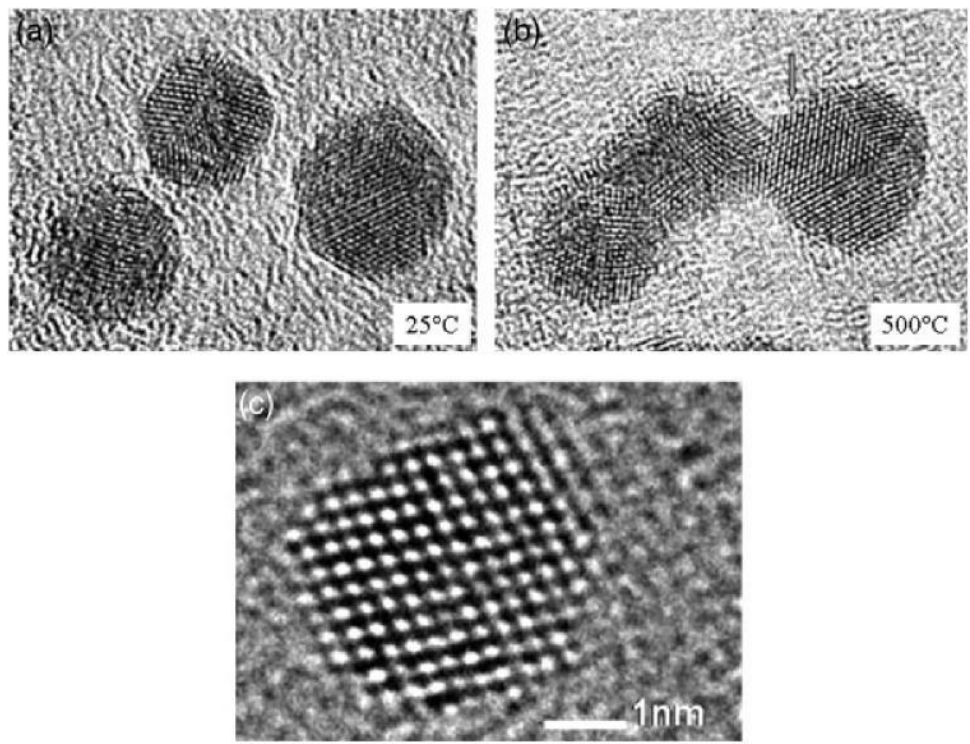

Fig. 12. Dynamic in situ studies of gold nanoparticle sintering (a) unheated (25°C) and (b) at 500°C, with alignment of (111) crystal planes (0.23 nm) during migration and sintering (2AC-TEM/STEM).

(c) Image detail of performance critical size range of gold nanoparticle at 25°C showing both internal perfection and (111) surface terracing with discrete steps (in 2AC TEM/STEM).

crystal structure of gold nanoparticles and sintering behavior were not reported (Chen and Goodman, 2004). The sintering studies presented here have implications in understanding the nucleation and subsequent growth of nanoparticles, metal nanowires, and nanorods. Recent experiments in a 1A double aberration corrected environment with STEM and TEM show insights into strong metal-support interactions of supported antimicrobial nanoclusters (Buckley et al., 2008).

## CONCLUSIONS

Dynamic in situ experiments can be very efficient, and productive of data, in scanning a range of experimental conditions in a single session, saving time and costs. As presented here, we are currently developing aberration corrected (AC) TEM and AC-STEM for dynamic in situ experiments. This is important in promoting a CTF extending to higher frequencies and spatial resolutions, allowing image recording at close to zero defocus. Recent dynamic in situ experiments in a 1A double aberration corrected environment show that dynamic analyses of reacting ultra small (<2 nm) nanoparticles and clusters on supports, including strong metal support interactions, are becoming possible. The mechanisms of change are revealed with new levels of image detail in dynamic in situ experiments under controlled conditions using a regular hot stage at close to 1A resolution.

## ACKNOWLEDGMENTS

The authors thank L. Hanna (DuPont) for the assistance in developing ETEM. They also thank the University of York, regional development agency, Yorkshire Forward, the European Union through the European Regional Development Fund and JEOL (UK) Ltd., for the Nanocenter facilities. Ian Wright is acknowledged for assistance in sample preparation.

## REFERENCES

Allen LJ, Findlay SD, Lupini AR, Oxley MP, Pennycook SJ. 2003. Atomic resolution EELS imaging in AC-STEM. Phys Rev Lett 91:105503.
Anderson JR. 1975. Structure of metallic catalysts. London: Academic. p. 1.
Batson P, Dellby N, Krivanek O. 2002. Sub-Angstrom resolution using AC-electro optics. Nature 418:617–619.
Bondize V, Parker SC, Campbell CT. 1999. Studies of gold particles. Catal Lett 63:143.
Boyes ED, Gai PL. 1997. Environmental high resolution electron microscopy and applications to chemical science. Ultramicroscopy 67:219–229.
Buckley J, Gai PL, Lee A, Wilson K. 2008. Silver carbonate nanoparticles over alumina stabilised nano-needles exhibiting potent antibacterial properties. Chem Commun 4013–4015.
Butler EP, Hale KF. 1981. Dynamic experiments. Amsterdam: North Holland. p. 10.
Chen M, Goodman DW. 2004. The structure of chemically active gold. Science 306:252–254.
Cottrell AH. 1971. Introduction to metallurgy. United Kingdom: Arnold. p. 1.
Crozier P, Sharma R, Datye A. 1998. Oxidation and reduction of small Pd particles on silica. Microsc Microan 4:278–279.
Ferreira PJ, Mitsuishi K, Stach E. 2008. In situ TEM. MRS Bull 33:83–85.

Gai PL. 1981. Dynamic studies of metal oxide catalysts MoO$_3$. Philos Mag 43:841–858.

Gai PL, McCarron E. 1990. Direct observation of CuO$_2$ shear defects in L$_{a2-xSr_x}$ CuO$_4$. Science 247:553–555.

Gai PL, Smith BC. 1990. In situ ETEM studies of intermetallic compound catalysts. Ultramicroscopy. 34:17–27.

Gai PL. 1992. Defects in oxide catalysts: Fundamental studies of catalysis inaction. Cat Rev Sci Eng 34:1–54.

Gai PL, et al. 1995. Solid state defect mechanism in vanadyl pyrophophate: Implications for selective oxidation. Science 267:661–663. Gai PL. 1997. A new transformation mechanism in catalytic oxides. Acta Crystallogr B53:346–352.

Gai PL. 1998. Direct imaging of gas molecule-solid catalyst interactions on the atomic scale. Adv Mat 10:1259–1263.

Gai PL. 2006. Pioneering Development of Environmental E(S)TEM. Electron Microscopy, Proc. International Microscopy Cong., IMC16, Sapporo, Japan, 1:53–54.

Gai PL. 2007. In situ environmental transmission electron microscopy. Nanocharacterisation In: Kirkland A, Hutchison JL, editors. Royal Society of Chemistry, Ch 7. p. 268–290.

Gai PL, Boyes ED. 2003. Electron microscopy in heterogeneous catalysis. Bristol (UK) and Philadelphia (USA): Institute of Physics Publishing. p. 176.

Gai PL, Boyes ED. 2005. Atomic resolution in situ ETEM studies: Microsc Microan 11:1526–1527.

Gai PL, Boyes ED. 2008a. A very high temperature (2000°C) stage for atomic resolution in-situ ETEM. Proceedings of 14th European Microscopy Congress. In: Luysberg M, Tillman K, Weirich T, editors. Instrumentation and methods, Vol. 1. Aachen, Germany: Springer-Berlin. p. 481–482.

Gai PL, Boyes ED. 2008b. Dynamic in situ experiments in a 1A° double aberration corrected environment. Proceedings of 14th European Microscopy Congress. In: Luysberg M, Tillman K, Weirich T, editors. Instrumentation and methods, Vol. 1. p. 479.

Gai PL, Boyes ED, Hansen P, Helveg S, Giorgio S, Henry C. 2007a. Atomic scale. Mater Res Soc Bull 32:1044–1048.

Gai PL, Torardi CC, Boyes ED. 2007b. Turning points in solid state chemistry. In: Harris KDM, Edwards PP, editors. For Sir John Meurig Thomas 75th birthday symposium. Ch. 45. London: Royal Society of Chemistry. p. 745.

Gai PL, Sharma R, Ross FM. 2008. Environmental (S)TEM studies of gas-liquid-solid interactions under reaction conditions. Mater Res Soc Bull 33:107–110.

Gall K, et al. 2005. Stability of nanoclusters. Mater Res Soc Symp Proc 854E:U5.7.

Gmelin. 1993. Handbook of inorganic and organometallic chemistry, Tungsten (W) system. Supp. A5b. Heidelberg: Springer-Verlag.

Haggin J. 1995. Catalysis at the atomic level. Am Chem Soc C&E News 73:39–42.

Haider M, et al. 1998. Electron microscopy image enhanced. Nature 392:768–769.

Hansen TW, Wagner J, Hansen PL, Dahl S, Topo SH, Jacobson J. 2001. Science 294.

Isaacson M, Kopf D, Uylaut M, Parker NW, Crewe AV. 1977. Direct observation of atomic diffusion by STEM. Proc Natl Acad Sci 74:1802–1806.

Hirsch PB, Howie A, Nicholson R, Pashley DW, Whelan MJ. 1985. Electron microscopy of thin crystals. New York: Kruger Publishing.

Jacoby M. 2002. Am Chem Soc C&E news 80:26–29.

Kamino T, Saka H. 1993. Microsc Microanal Microstruct 4:127.

Kan C, Wang G, et al. 2006. Appl Phys Lett 88:071904.

Marks LD, Smith DJ. 1981. HREM studies of small particles. J Cryst Growth 54:425–430.

Massalski TD. 1986. Binary alloy phase diagrams. Am Soc Met 1– 2:599–560.

Mohamed MB, Wang Z, El Sayed MA. 1999. Temp-dependent size controlled growth of gold nanoclusters. J Phys Chem 103:10255–10259.

Oyama ST. 1996. Chemistry of transition metal carbides. London: Blackie Academic and Professional.

Sinclair R, Yamashita T, Ponce F. 1981. In situ HREM. Nature 290:386–388.

Sharma R, Rez P, Brown M, Du GH, Treacy M. 2007. Carbon nano-tubes. Nanotechnology 18:125602.

Thomas JM, Terasaki O, Gai PL, Zhou W, Gonzalez-Calbet. 2001. Acc Chem Res 34:583

Thomas JM, Gai PL. 2004. Electron microscopy and materials chemistry of solid catalysts. Adv Catal 48:171–227.

Yen C, Shimizu K, Lin Y, Bailey F. 2007. Pt-based bimetallic nanoparticles for direct methanol fuel cells. Energy Fuels (ACS) 21:2268–2271.

Varykhalov A, Rader O, Gudat W. 2008. Au on surface WC. Phys Rev B 77:035412.

#www.Maxsidorov.com/ctfexplorer/index.htm.